\documentclass[aps,%
12pt,%
final,%
oneside,
onecolumn,%
musixtex, %
superscriptaddress,%
centertags]{article} %%
\usepackage[utf8]{inputenc}
\usepackage[T2A]{fontenc} % Поддержка русских букв
\usepackage[colorlinks=true,linkcolor=blue,unicode=true]{hyperref}
\usepackage{euscript}
\usepackage{supertabular}
\usepackage[pdftex]{graphicx}%
\usepackage{amsthm,amssymb, amsmath}
\usepackage{textcomp}
\usepackage[noend]{algorithmic}
\usepackage[ruled]{algorithm}
\usepackage{braket}
\usepackage{amsmath}
\usepackage{comment}
\usepackage{setspace}
 \DeclareMathOperator{\Tr}{Tr}
\usepackage{titlesec}
\usepackage{setspace}

\usepackage[numbers,sort&compress]{natbib}

\usepackage{hyperref}
\hypersetup{colorlinks,
citecolor=blue
}

\newcommand{\be}{\begin{equation}}
\newcommand{\ee}{\end{equation}}
\newcommand{\ba}{\begin{eqnarray}}
\newcommand{\ea}{\end{eqnarray}}
\newcommand{\bc}{\begin{comment}}
\newcommand{\ec}{\end{comment}}
\newcommand{\eps}{\varepsilon}

\usepackage{amsthm}

\begin{document}
%\pdfrender{StrokeColor=black,TextRenderingMode=2,LineWidth=0.5pt}

\begin{center}

{\Large\bf A perturbation algorithm for the pointers of Franke-Gorini-Kossakowski-Lindblad-Sudarshan~equation}

\vspace{1cm}
{\large A. A. Andrianov$^{1,2,}$\footnote{E-mail: a.andrianov@spbu.ru}, M. V. Iof\/fe$^{1 ,}$\footnote{E-mail: m.ioffe@spbu.ru},
E. A. Izotova$^{1,3,4,}$\footnote{E-mail: ekat.izotova@gmail.com}, O.~O.~Novikov$^{1 ,}$\footnote{E-mail: o.novikov@spbu.ru}}
\\
\vspace{0.5cm}
$^1$ Saint Petersburg State University, 7/9 Universitetskaya nab., St. Petersburg, Russia 199034\\
$^2$ Departament de Física Quàntica i Astrofísica and Institut de Ciències del Cosmos (ICCUB), Universitat de
Barcelona, Martí i Franquès 1, 08028 Barcelona, Spain\\
$^3$ Skolkovo Institute of Science and Technology, Bolshoy Boulevard 30, bld. 1, Moscow 121205, Russia\\
$^4$ Moscow Institute of Physics and Technology, Institutsky lane 9, Dolgoprudny, Moscow Region, Russia
141700
\end{center}

\vspace{1cm}

\begin{abstract}
This paper is devoted to the study of behavior of open quantum systems consistently
based on the Franke–Gorini–Kossakowski–Lindblad–\\Sudarshan (FGKLS) equation
which covers evolution in situations when decoherence can be distinguished. We focus on
the quantum measurement operation which is determined by final stationary states of an open
system—so called pointers. We find pointers by applying the FGKLS equation to asymptotically
constant density matrix. In seeking pointers, we have been able to propose a perturbative
scheme of calculation, if we take the interaction components with an environment to
be weak. Thus, the Lindblad operators can be used in some way as expansion parameters
for perturbation theory. The scheme we propose is different for the cases of non-degenerate
and degenerate Hamiltonian. We illustrate our scheme by particular examples of quantum
harmonic oscillator with spin in external magnetic field. The efficiency of the perturbation
algorithm is demonstrated by its comparison with the exact solution.
\end{abstract}


%\UseRawInputEncoding
%\tableofcontents
%\newpage

\section{Introduction}
In the last decade, the interest to evolution of open quantum systems was growing with the
development of quantum information technology. The evolution is accompanied by dissipation
of quantum states and by decoherence, the process, in which pure quantum states
transform into mixed ones, that contain classical probabilities, beside quantum ones. The
quantum information contained in an initial state is getting partially lost in the environment.
In studying such systems, one may progress further in understanding of how the quantum
world becomes the classical one we live in. This quantum-to-classical transition is widely discussed in the literature, see, for example, the books \cite{breuer,schlosshauer,weiss}, references therein, and also more recent papers \cite{PNAS,quantum}.

The examination of behavior of open quantum systems is based on the equation which
covers a dynamical semigroup evolution when decoherence can be distinguished,
\be
\label{lindblad}\dot{\rho} = -i[H,\rho]+\sum_a L^{(a)}  \rho L^{(a),\dag} - \frac{1}{2} \left\{\sum_a  L^{(a),\dag} L^{(a)},\rho\right\}
\ee
where $\rho$ is a density matrix of the system under study, $\{L^{(a)}\}$ is a set of operators carrying information about an interaction of the system with an environment. $\left\{\right\}$ denotes an anticommutator.
%ДОБАВИТЬ ССЫЛКУ%%%%%%%%%%%%%%%%%%%%%%%%%%%%%%%%%%%%%

This equation was obtained independently by Lindblad \cite{lindblad1, lindblad2} and Franke \cite{franke} as well as by
Gorini et al. \cite{gorini} and we name it further on as the FGKLS equation.

For this exact form of equation, there are several premises. When focusing on nondissipative
processes we assume that the density matrix in evolution in time must remain
Hermitian, positive, and its trace must be equal to $1$. It implicates that density matrix always
sticks to its definition: it contains in itself a notion of non-negative probabilities of the realization
of some pure quantum states and an assumption that the entire probability stays the
same. The opposite of the latter would mean a decay of the system or some other event of
the leak of the total probability. However if the dissipation holds one could factor it out to
proceed to conditional probabilities for conditional density matrices \cite{andrianovtarrach}. The reduced equation
becomes nonlinear. Such situations are not considered in this work. If all the properties
mentioned above are preserved, and we require the master equation to be linear, providing
the state superposition, then the only possible form of equation is the FGKLG form (for this
statement a more strict property—complete positivity—is necessary \cite{benatti_floreanini,bf-2}).

One could also consider the composite system (system of interest + environment) to
obey the Liouville law of evolution in time with a Hamiltonian $H$ (usually of the form $H_S + H_E + H_{int}$)
associated to the whole system . Then after tracing out the environmental
degrees of freedom the density matrix of the space $\mathcal{H}_S \otimes \mathcal{H}_E$ is reduced to the density matrix on the subspace $\mathcal{H}_S$
and in the weak coupling limit the equation approaches the FGKLS
form \cite{benatti_floreanini,bf-2}. An example of open quantum system is one or a pair of optically active noninteracting
atoms (or molecules) weakly coupled to a heat bath. This model was studied in
\cite{benatti_floreanini,bf-2}.

A variety of application areas exists for the FGKLS equation: from solid state physics
and quantum optics \cite{breuer} to high energy physics \cite{Blaizot, hep}, and even in the search for quantum
mechanics breaking in vacuum due to tiny tracks of quantum gravity (see review in \cite{qgrav}).

The study of open quantum systems is related in particular to the problem of quantum
measurements: a measuring apparatus can be considered as an environment that interferes
into a quantum state of the system—particle, molecule, other (see classical papers of Landau
\cite{landau} and von Neumann \cite{neumann}, more recent papers \cite{zurek,z-2} and the textbook \cite{weinberg_2016}). As a result,
a system evolves, constantly interacting with an environment, in such a way as to produce
the result of a measurement. An appropriate description of decoherence of a state of an open
system turns out to be the density matrix instead of a usual state vector, because mixed states
cannot be discussed in any other way. But, still, the standard approach is understanding the
mixed state as a composition of state vectors with various classical probabilities. Recently,
S.Weinberg also made his point, that we should give up the state language and consider the
density matrix as a primary mathematical object to describe quantum world \cite{weinberg_2014}.

The measurement operation is determined by final stationary states of an open system — so
called \textbf{pointers}. We expect them to be {\it steady,} as they represent the readings of a classical
instrument. In the present paper, we entirely focus on the small decoherence with $\|L^{(a)}\|^2 \ll \|H\|$,
when the contribution of Lindblad part of the equation is much smaller than that of the
Hamiltonian part. We find pointers by applying to the FGKLS equation (\ref{lindblad}) an extra condition:
\be
\label{statpointer} \dot\rho = 0.
\ee
The term “pointer” was introduced by W.H. Zurek in above-mentioned papers \cite{zurek,z-2}. In
this manner, he called the environment-superselected preferred states. The idea was to relate
them to a measuring device.

The organization of this paper is as follows. In Sect. \ref{sec2}, the notations are given, and
the main equations are prepared in a suitable form using the energy eigenstate basis. In
Sect. \ref{sec3}, in seeking pointers, we propose a scheme of calculation taking the interaction with
an environment to be weak, $\|L^{(a)}\|^2 \ll \|H\|$. Thus, the jump operators $L^{(a)}$ can be used
in some way as expansion parameters for perturbation theory. The scheme we propose is
different for the cases of non-degenerate and degenerate Hamiltonian in (\ref{lindblad}). The degeneracy
of the Hamiltonian spectra may arise either as a consequence of global symmetry of the system
or due to a (quasi) level crossing. The perturbation algorithm for a degenerate spectrum is
presented in Sect. \ref{sec32}.

A few considerations and attempts to elaborate the perturbation expansion for the FGKLS
equation have been undertaken in \cite{BNN,PT1,PT2,PT3} at a more symbolic form using a deviation from
the bare Hamiltonian taken as a small expansion parameter. As compared with them our
expansion treats the operators generating decoherence as perturbations and as well includes
the option of (almost) degenerate energy levels when the jump coefficients break the Hamiltonian
degeneracy and tune respective energy eigenstates in a particular direction in the
Hilbert space: interaction with an environment directs the system toward a set of steady
states, pointers.

In Sect. \ref{sec4} we study particular examples in order to illustrate our scheme of calculation.
They include a number of models of quantum harmonic oscillator with spin interacting with
external magnetic field. The efficiency of the perturbation algorithm is demonstrated by its
comparison with the exact solution. In Conclusions, a summary of our results is given and
possible directions for future research are outlined.

\section{The first steps}\label{sec2}

Planning to develop the perturbation approach to construction of asymptotically steady
states — pointers of the FGKLS equation  (\ref{lindblad}) for the density matrix $\rho,$ we start from the
solution  $\rho_0$ of the Liouville–von Neumann equation:
\be
\dot{\rho_0} = -i[H,\rho_0] = 0, \label{liouville}
\ee
with some Hermitian Hamiltonian $H$ which does not depend on time. Further on, we suppose
that the Lindblad operators $L^{(a)}$ in (\ref{lindblad}) are small, in comparison with Hamiltonian $H$, $\|L^{(a)}\|^2 \ll \|H\|$,
and they serve as expansion parameters of perturbation theory. For brevity
(as long as it does not change the content) in all formulas below, we will assume that we are
dealing with a single Lindblad operator $L.$ The appearance of additional Lindblad operators
$L^{(a)},\, a=1,2,\dots $, will lead to an extra summation over the index $a$.

Thus, we take
\be
\rho = \rho_0 + \Delta \rho \label{rr}
\ee
where $\Delta \rho$ is presumed to be small, compared to $\rho_0$: $\|\Delta\rho\| \ll \|\rho_0\|$.

The next step is to decompose all the operators in the energy basis: the orthonormal
eigenvectors $\{\ket{E_k}\}$ of the Hamiltonian $H:$
\be \label{h} H=\sum_k \eps_k \ket{E_k} \bra{E_k}, \ee
where both eigenvalues $\{\eps_k\}$ and state vectors $\{\ket{E_k}\}$ do not depend on $t.$

We shall use the following notations for expansions:
\ba
L &=& \sum_{ij} l_{ij} \ket{E_i} \bra{E_j}; \label{l} \\
\rho &=& \sum_{ij} f_{ij}(t) \ket{E_i} \bra{E_j}; \label{rho} \\
\rho_0 &=& \sum_{ij} g_{ij}(t) \ket{E_i} \bra{E_j} \label{rho0_0}
\ea
with a usual condition for a density matrix:
\be
\Tr \rho = \Tr \rho_0 = 1. \label{trace}
\ee

The coefficients $\{l_{ij}\}$ of (\ref{l}) will be our perturbation theory expansion parameters. It seems
to be evident that $\Delta \rho$ must be small compared to $\rho_0$.. Namely, the expansion for $\rho_0$ in terms
of $l_{ij}$ should start with the zeroth-order terms $\sim l^0$ (by $l$ we denote any matrix element $l_{ij}$),
while the expansion for $\Delta \rho$ should start with the higher order, i.e., $\sim l^k, k>0$. Nonetheless,
it is not the case. As we will see, it is not possible to fix $\rho_0$ so that we could choose such small
$\Delta \rho$ for $\rho_0 + \Delta \rho$ to satisfy Eq. (\ref{lindblad}) with the ‘pointer’ condition (\ref{statpointer}). It does not matter what $\rho_0$
we had in Liouville case, after turning on the FGKLS interaction, the resulting pointer state
$\rho$ would become exactly one obeying certain equations. This is the reason why we seek $\rho$,
not $\Delta \rho$, and after finding $\rho$ we compare it to $\rho_0$ to see if they are close.

Substituting now (\ref{h}), (\ref{l}), (\ref{rho}) into (\ref{lindblad}) (and noting that $\{\ket{E_k}\}$ is a stationary basis) we get for the pointer (\ref{statpointer}):
\be\label{allmn}
-i f_{mn}(\eps_m-\eps_n) + \sum_{k,l}l_{mk}f_{kl}l^{*}_{nl} - \frac{1}{2}\sum_{k,l}l^{*}_{km}l_{kl}f_{ln}-\frac{1}{2}\sum_{k,l}f_{mk}l^{*}_{lk}l_{ln} = 0
\ee
for all $m$ and $n$. Here, $\eps_n$ has the zeroth order, $l_{ij}$ are small expansion parameters, and $f_{mn}$ are amenable to expansion in a power series in $l_{ij}$.

Below, two cases will be considered separately. Namely, the case $m=n$ with the first
term in (\ref{allmn}) vanishing
\be
\sum_{k,l}l_{mk}f_{kl}l^{*}_{ml} - \frac{1}{2}\sum_{k,l}l^{*}_{km}l_{kl}f_{lm} -\frac{1}{2}\sum_{k,l}f_{mk}l^{*}_{lk}l_{lm} = 0 \label{mm}
\ee
will be called as \textbf{Equation 1}, and the case $m\neq n,$ for which the corresponding Eq. (\ref{allmn}) will
be called as \textbf{Equation 2}. The perturbation expansion differs crucially for these two cases.

By the substitution of the expansion (\ref{rho0_0}) into the Eq. (\ref{liouville}), one obtains the Liouville pointers:
\be
\dot{g}_{mn} = -i g_{mn}(\eps_m-\eps_n) = 0, \quad t\to\infty, \label{mnmn}
\ee
for arbitrary indices $m$ and $n$.
This means that asymptotically the diagonal elements $g_{mm}(t)$ are arbitrary constants, while for off-diagonal $m\neq n$ we have to consider two options:
\begin{itemize}
\item $H$ is non-degenerate, i.e. for $m\neq n,$ $\eps_m \neq \eps_n$ and asymptotically $g_{mn}(t) \rightarrow 0.$ Thus, the off-diagonal elements of density matrices $\rho_0$ for the Liouville pointers vanish at large $t.$
\item $H$ is degenerate, i.e. there are some different states $\ket{E_m}$ and $\ket{E_n}$ ($m\neq n$) for which $\eps_m=\eps_n.$ From (\ref{mnmn}) it is clear that in this case, $g_{mn}(t)$ are asymptotically arbitrary constants both for such indices $m,\, n,$ and for diagonal elements with $m=n$, while $g_{mn}(t)\rightarrow 0$ for all other pairs of $m,\,n.$

\end{itemize}

\section{The perturbation scheme of calculation of FGKLS pointers}\label{sec3}

To start with solving  (\ref{lindblad}) perturbatively we replace the Lindblad operator as follows:
\be
L \to \lambda L;\quad l_{ij} \to \lambda l_{ij}; \quad \lambda \ll 1.
\label{lambda}
\ee
Now, we suppose that $f_{mn}$ have a decomposition
\be
f_{mn} = f_{mn}^{(0)} + f_{mn}^{(1)} + \ldots; \quad  f_{mn}^{(k)} =  c_k \;\lambda^{2k} \label{exp}
\ee
where $c_k$ are c-numbers which include dependence on finite values of $l_{ij}.$
We expect the expansion in even powers $\lambda^{2k}$, because \textbf{Equations 1} and \textbf{2} are quadratic in small coefficients $\lambda.$
Thus, eventually the expansion has the form:
\be
f_{mn} = f_{mn}^{(0)} + f_{mn}^{(1)} + f_{mn}^{(2)} +\ldots = c_0 \lambda^0 + c_1 \lambda^2 + c_2 \lambda^4 + \dots . \label{expp}
\ee
%Here $c_0, c_1, c_2,\dots$ have a conventional character, since the term $c_2 l_4$, for example, contains terms with different fourth order products of various $l_{ij}$ with some coefficients.
The trace restriction (\ref{trace}) provides a set of conditions:
\be
\sum_m f^{(0)}_{mm} = 1; \quad  \sum_m f^{(k)}_{mm} = 0,\,\, k=1, 2 \dots   \label{trace+}
\ee

The above-mentioned \textbf{Equations 1} and \textbf{2} include an arbitrary fixed-order $s$ of perturbation expansion for coefficients $f^{(s)}_{ij}$ as follows:\\
\textbf{Equation 1}:
\ba
&&\sum_i |l_{mi}|^2 f^{(s)}_{ii} - \sum_i |l_{im}|^2 f^{(s)}_{mm} + \sum_{i,j;i\neq j}l_{mi}f^{(s)}_{ij}l^{*}_{mj} \nonumber\\
&&-\frac{1}{2}\sum_{i,j;j\neq m}l^{*}_{im}l_{ij}f^{(s)}_{jm}-\frac{1}{2}\sum_{i,j;j\neq m}f^{(s)}_{mj}l^{*}_{ij}l_{im} = 0 \label{main1}
\ea
for an arbitrary fixed index $m$.\\
\textbf{Equation 2}:
\ba
&&-i f^{(s)}_{mn}(\eps_m-\eps_n) + \lambda^2\sum_i l_{mi}f^{(s-1)}_{ii} l^{*}_{ni} - \frac{1}{2}\lambda^2\sum_i l^{*}_{im}l_{in}(f^{(s-1)}_{mm}+f^{(s-1)}_{nn}) \nonumber\\
&& + \lambda^2\sum_{i,j;i\neq j}l_{mi}f^{(s-1)}_{ij}l^{*}_{nj} - \frac{1}{2}\lambda^2\sum_{i,j;j\neq n}l^{*}_{im}l_{ij}f^{(s-1)}_{jn}\nonumber\\
&&-\frac{1}{2}\lambda^2\sum_{i,j;j\neq m}f^{(s-1)}_{mj}l^{*}_{ij}l_{in} = 0,    \label{main2}
\ea
for arbitrary fixed pair $m\neq n.$

Now we have to consider a Hamiltonian of non-degenerate and degenerate kinds separately, since in the latter case the first term in \textbf{Equation 2} (\ref{main2}) disappears for $\eps_n = \eps_m,$ substantially changing an application of the perturbation theory.

\subsection{Non-degenerate Hamiltonian}
\begin{itemize}
\item
As the first step, we shall rewrite \textbf{Equation 1} (\ref{main1}) by gathering the diagonal elements  $f^{(s)}_{mm}$ in the l.h.s., but the non-diagonal elements $f^{(s)}_{mn}, m\neq n$ of the same order $s$ in the r.h.s. In the matrix form, the result is:
\ba
&&\begin{pmatrix}
-\sum_{t\neq 1}|l_{t1}|^2 & |l_{12}|^2 & |l_{13}|^2 & \dots \\
|l_{21}|^2 & -\sum_{t\neq 2}|l_{t2}|^2 & |l_{23}|^2 & \dots\\
|l_{31}|^2 & |l_{32}|^2 & -\sum_{t\neq 3}|l_{t3}|^2 & \dots \\
\vdots & \vdots & \vdots & \ddots \\
\end{pmatrix}
\begin{pmatrix}
f^{(s)}_{11} \\
f^{(s)}_{22} \\
f^{(s)}_{33} \\
\vdots \\
\end{pmatrix}
 \,\,\label{nondeg1}\nonumber \\
&&=\begin{pmatrix}
-\sum_{i,j;i\neq j}l_{1i}f^{(s)}_{ij}l^*_{1j} + \frac{1}{2}\sum_{i,j;j\neq 1}l^*_{i1}l_{ij}f^{(s)}_{j1}+\frac{1}{2}\sum_{i,j;j\neq 1}f^{(s)}_{1j}l^*_{ij}l_{i1} \\
-\sum_{i,j;i\neq l}l_{2i}f^{(s)}_{ij}l^*_{2j} + \frac{1}{2}\sum_{i,j;j\neq 2}l^*_{i2}l_{ij}f^{(s)}_{j2}+\frac{1}{2}\sum_{i,j;j\neq 2}f^{(s)}_{2j}l^*_{ij}l_{i2}  \\
-\sum_{i,j;i\neq j}l_{3i}f^{(s)}_{ij}l^*_{3j} + \frac{1}{2}\sum_{i,j;j\neq 3}l^*_{i3}l_{ij}f^{(s)}_{j3}+\frac{1}{2}\sum_{i,j;j\neq 3}f^{(s)}_{3j}l^*_{ij}l_{i3}  \\
\vdots \\
\end{pmatrix}. \nonumber \\ 
&& 
\ea
The existence of solutions $f_{mm}^{(s)}$ to this system of equations depends on the properties of the matrix in the l.h.s. and on the column in the r.h.s.. Summing up all the lines in the matrix, we get the null line, i.e. its determinant vanishes. In the case of homogeneous system (with zero column in the r.h.s.), the solution can be found, and it contains one or more free parameters depending on the rank of the matrix. One free parameter is just necessary to fulfill the trace condition (\ref{trace}). In the case of inhomogeneous system (with nonzero column in the r.h.s.), according to Kronecker–Capelli theorem, a solution of system of equations (\ref{nondeg1}) exists if the rank of the matrix is equal to the rank of the augmented matrix (the matrix with the right part of the equation glued to it). A number of parameters in the solution also depends on the rank of the matrix, but, anyway, it is more than $0$, so the trace condition (\ref{trace}) can be imposed. If the mentioned ranks are not equal to each other, then there is no solution. In such a case, our scheme would not work at all.
But we have a sign that the solution of (\ref{nondeg1}) actually can exist. One can easily check, that the sum of all the equations leads to identity $0=0.$ So, the system of equations is consistent, at least.

\item
The second step is finding non-diagonal elements $f^{(s)}_{mn}$ from \textbf{Equation 2} (\ref{main2}), if we know diagonal $f^{(s-1)}_{mm}$ and non-diagonal $f^{(s-1)}_{mn}, m\neq n$ elements of the previous order $s-1.$ Let us rewrite \textbf{Equation 2} (\ref{main2}) for $m\neq n$ as:
\ba
&&f^{(s)}_{mn} = \frac{i\lambda^2}{\eps_m-\eps_n}\Big[ -\sum_i l_{mi}f^{(s-1)}_{ii} l^{*}_{ni} + \frac{1}{2}\sum_i l^{*}_{im}l_{in}(f^{(s-1)}_{mm}+f^{(s-1)}_{nn})  \nonumber\\
&&- \sum_{i,j;i\neq j}l_{mi}f^{(s-1)}_{ij}l^{*}_{nj} + \frac{1}{2}\sum_{i,j;j\neq n}l^{*}_{im}l_{ij}f^{(s-1)}_{jn}+
\frac{1}{2}\sum_{i,j;j\neq m}f^{(s-1)}_{mj}l^{*}_{ij}l_{in}\Big] \label{nondeg2}
\ea

\end{itemize}

Our scheme of calculations is as follows:
\begin{itemize}
\item To start with, due to the obvious fact that $f^{(-1)}_{mn} \equiv 0$ for $m\neq n$, one obtains from Eq. (\ref{nondeg2}) that in the zeroth order
    $f^{(0)}_{mn} = 0$ for $ m\neq n.$
\item Then, we substitute $f^{(0)}_{mn}=0, m\neq n$ into Eq. (\ref{nondeg1}). The column in the r.h.s. vanishes, and we get a homogeneous equation with degenerate matrix. Therefore the solution $\{f^{(0)}_{mm}\}$ exists with one or more free parameters, and after that, one parameter is eliminated by the trace condition
    $\sum_m f^{(0)}_{mm}=1.$
\item We put the zeroth order $f^{(0)}_{mn}=0, m\neq n$ and $f^{(0)}_{mm}$ obtained above into Eq. (\ref{nondeg2}) to find the first order nondiagonal $f^{(1)}_{mn}, m\neq n.$
\item We substitute $f^{(1)}_{mn}, m\neq n$ into Eq. (\ref{nondeg1}). According to Kronecker–Capelli theorem, we find either the solution for diagonal $f^{(1)}_{mm}$ with one or more parameters, or no solutions. In the former case, one parameter of all is eliminated by the trace condition $\sum_m f^{(1)}_{mm} = 0.$
\item If the previous step is successful, we find $f^{(2)}_{mn}, m \neq n$ from Eq. (\ref{nondeg2}) by substitution of $f^{(1)}_{mn}, m\neq n$ and $f^{(1)}_{mm}.$
\item And so on.
\end{itemize}

At the steps of this scheme, in which we must apply Kronecker–Capelli theorem, there is a probability that the scheme crashes down (if the theorem states that there is no solution). %That is an open problem that needs further investigation.

We must also note here that the FGKLS pointers have an evident difference from Liouville pointers: their diagonal elements are already not arbitrary at the zeroth order. After "turning Lindblad on", the density matrix changes essentially - it chooses a direction in which it "goes". And conversely, after slowly "turning Lindblad off" (i.e., taking the solution for the Lindblad pointers $\rho$, setting all the terms with $L$ to vanish) some "traces" of Lindblad equation remain.

\subsection{Degenerate Hamiltonian}\label{sec32}
In this case, more delicate study has to be implemented. \textbf{Equation 2} (\ref{main2}) differs for $(m,n)$ such that $\eps_m=\eps_n$ (let us call this pair $(m,n)$ "internal") and $(m,n)$ for which $\eps_m\neq\eps_n$ (we will call this pair $(m,n)$ "external"). Separating $\{f^{(s)}_{mn}\}$ with internal and external indices, we write \textbf{Equation 1} (\ref{main1}) and the two versions of \textbf{Equation 2} (\ref{main2}) in the following form:\\
\textbf{Equation 1}:
\ba
&&\sum_i |l_{mi}|^2 f^{(s)}_{ii} - \sum_i |l_{im}|^2 f^{(s)}_{mm} + \sum_{\substack{i,j;i\neq j;\\(i,j)-\text{int}}}l_{mi}f^{(s)}_{ij}l^{*}_{mj} \nonumber\\
&&-\frac{1}{2}\sum_{\substack{i,j;j\neq m;\\(j,m)-\text{int}}}l^{*}_{im}l_{ij}f^{(s)}_{jm}-
\frac{1}{2}\sum_{\substack{i,j;j\neq m;\\(m,j)-\text{int}}}f^{(s)}_{mj}l^{*}_{ij}l_{im}  \nonumber\\
&&= - \sum_{\substack{i,j;i\neq j;\\(i,j)-\text{ext}}}l_{mi}f^{(s)}_{ij}l^{*}_{mj} +
\frac{1}{2}\sum_{\substack{i,j;j\neq m;\\(j,m)-\text{ext}}}l^{*}_{im}l_{ij}f^{(s)}_{jm}+
\frac{1}{2}\sum_{\substack{i,j;j\neq m;\\(m,j)-\text{ext}}}f^{(s)}_{mj}l^{*}_{ij}l_{im}
\label{deg1}
\ea
for some $m$.\\
\textbf{Equation 2}:
\ba
&&\sum_i l_{mi}f^{(s)}_{ii} l^{*}_{ni} - \frac{1}{2}\sum_i l^{*}_{im}l_{in}(f^{(s)}_{mm}+f^{(s)}_{nn})  \nonumber\\
&&+ \sum_{\substack{i,j;i\neq j;\\(i,j)-\text{int}}}l_{mi}f^{(s)}_{ij}l^{*}_{nj} -
\frac{1}{2}\sum_{\substack{i,j;j\neq n;\\(j,n)-\text{int}}}l^{*}_{im}l_{ij}f^{(s)}_{jn}-
\frac{1}{2}\sum_{\substack{i,j;j\neq m;\\(m,j)-\text{int}}}f^{(s)}_{mj}l^{*}_{ij}l_{in}  \nonumber\\
&&= - \sum_{\substack{i,j;i\neq j;\\(i,j)-\text{ext}}}l_{mi}f^{(s)}_{ij}l^{*}_{nj} +
\frac{1}{2}\sum_{\substack{i,j;j\neq n;\\(j,n)-\text{ext}}}l^{*}_{im}l_{ij}f^{(s)}_{jn}+
\frac{1}{2}\sum_{\substack{i,j;j\neq m;\\(m,j)-\text{ext}}}f^{(s)}_{mj}l^{*}_{ij}l_{in} \label{deg21}
\ea
for internal $(m,n)$, $m\neq n$.

\ba
&&\!\!\!\! f^{(s)}_{mn}=\frac{i\lambda^2}{\eps_m-\eps_n}\Big[ - \sum_i l_{mi}f^{(s-1)}_{ii} l^{*}_{ni} + \frac{1}{2}\sum_i l^{*}_{im}l_{in}(f^{(s-1)}_{mm}+f^{(s-1)}_{nn}) \nonumber\\
&&\!\!\!\! - \sum_{\substack{i,j;i\neq j;\\(i,j)-\text{int}}}l_{mi}f^{(s-1)}_{ij}l^{*}_{nj} +
\frac{1}{2}\sum_{\substack{i,j;j\neq n;\\(j,n)-\text{int}}}l^{*}_{im}l_{ij}f^{(s-1)}_{jn}+
\frac{1}{2}\sum_{\substack{i,j;j\neq m;\\(m,j)-\text{int}}}f^{(s-1)}_{mj}l^{*}_{ij}l_{in} \nonumber\\
&&\!\!\!\! - \sum_{\substack{i,j;i\neq j;\\(i,j)-\text{ext}}}l_{mi}f^{(s-1)}_{ij}l^{*}_{nj} +
\frac{1}{2}\sum_{\substack{i,j;j\neq n;\\(j,n)-\text{ext}}}l^{*}_{im}l_{ij}f^{(s-1)}_{jn}+
\frac{1}{2}\sum_{\substack{i,j;j\neq m;\\(m,j)-\text{ext}}}f^{(s-1)}_{mj}l^{*}_{ij}l_{in}\Big]    \label{deg22}
\ea
for external $(m,n)$, $m\neq n$.

The scheme of calculations is as follows:
\begin{itemize}
\item Analogously to non-degenerate case, due to $f^{(-1)}_{mn} \equiv 0$ for any $(m,n),$ one obtains from equation (\ref{deg22}) that in the zeroth order\\
    ${ f^{(0)}_{mn}=0, m\neq n, (m,n) - \text{external} }.$
\item Then, we substitute $\{f^{(0)}_{mn}=0, m\neq n, (m,n) - \text{external}\}$ into Eqs. (\ref{deg1}) and (\ref{deg21}). A composition of these two equations is a system of equations for $\{\{f^{(0)}_{mm}\}, { \{f^{(0)}_{mn}, m\neq n, (m,n) - \text{internal}\} } \}$ with vanishing r.h.s..
    %It means that the solution exists.
    If the determinant of the matrix corresponding to the system of equations is equal to $0$, then the solution has one or more free parameters. In such a case, we will then impose, as usual, the trace condition $\sum_m f^{(0)}_{mm} = 1,$ reducing the number of parameters by one. Otherwise, the solution has no parameters, and we are not able to impose the trace condition.
    %Then we substitute $\{f^{(0)}_{mn}=0, m\neq n, (m,n) - \text{external}\}$ into equations (\ref{deg1}) and (\ref{deg21}). A composition of these two equations is a system %of equations in $\{\{f^{(0)}_{mm}\}, { \{f^{(0)}_{mn}, m\neq n, (m,n) - \text{internal}\} } \}$ with the right part equal to $0$. It means that the solution exists. If %the determinant of the matrix corresponding to the system of equations is equal to $0$, then the solution has 1 or more parameters. This is desirable for us, since we %will then impose, us usual, the trace condition $\sum_m f^{(0)}_{mm} = 1$ and thus reduce the number of parameters by $1$. Otherwise, the solution has no parameters, and %we have a difficulty in imposing the trace condition.
\item We insert $\{f^{(0)}_{mn}\}$ with $(m,n)$ any possible pair of indexes (equal, internal or external) into Eq. (\ref{deg22}). Therefore, the r.h.s. is completely known, and we obtain ${ \{f^{(1)}_{mn}, m\neq n, (m,n) - \text{external}\} }.$
\item We substitute ${ \{f^{(1)}_{mn}, m\neq n, (m,n) - \text{external}\} } $ into Eqs. (\ref{deg1}) and (\ref{deg21}). We have inhomogeneous system of equations for\\
$\{\{f^{(1)}_{mm}\}, { \{f^{(1)}_{mn}, m\neq n, (m,n) - \text{internal}\} } \}$. Comparing the rank of the matrix corresponding to the system of equations and the rank of the augmented matrix and applying Kronecker–Capelli theorem, we again have a set of options. There are no solutions, or the solution doesn't have any parameters, or the solution has one or more free parameters. In the latter case, we can continue: impose the trace condition $\sum_m f^{(1)}_{mm} = 0,$ and to reduce the number of parameters by one.
\item We find $\{f^{(2)}_{mn}, m \neq n, (m,n) - \text{external}\}$ from Eq. (\ref{deg22}) (substituting the whole set $f^{(1)}_{mn}$, with $(m,n)$ equal, internal or external), if we succeed in finding $\{\{f^{(1)}_{mm}\}, { \{f^{(1)}_{mn}, m\neq n, (m,n) - \text{internal}\} } \}$ in the previous step.
\item And so on.
\end{itemize}

\section{Examples}\label{sec4}

Since for the case of a degenerate Hamiltonian matrix analysis of the system of linear equations is unclear to some extent, we will have a look at some examples.
The first two will be the models of a one-dimensional particle that moves in the potential of harmonic oscillator and possesses a spin interacting with external magnetic field. Interaction with the environment is described by one or two rather simple Lindblad operators, correspondingly. In the third example, the system with two-dimensional Hilbert space and one off-diagonal Lindblad operator will be considered.  As in the general discussion, our goal is to find pointers - asymptotically (at $t\to\infty$) stationary solutions of the Lindblad equation (\ref{lindblad}).

\subsection{First oscillator example}
To start with, we choose a model of one-dimensional harmonic oscillator in a constant magnetic field $B$ along $z-$axis:
\be
H = H_0 \otimes I + \delta\cdot I \otimes \sigma_3;\quad H_0 \equiv \frac{p^2}{2m} + \frac{1}{2}m \omega x^2; \quad \delta \equiv \frac{1}{2} \mu B \label{H}
\ee
and the single Lindblad operator $L$ in the form:
\be
L = \frac{1}{2}\lambda I \otimes \sigma_+; \quad \sigma_+ \equiv \sigma_1 + i \sigma_2, \label{L}
\ee
where $\lambda$ is a small $c-$number.

An appropriate energy basis $\ket{\Psi_M}$ will be a direct product of eigenvectors of $H_0$ and eigenvectors of $\sigma_3:$
\be
\{\ket{\psi_n}\otimes \ket{a}\},\,;n=0,1,\dots; a = 0,1;\,  H_0 \ket{\psi_n} = E_n \ket{\psi_n},  \,
E_n = \omega \left(n+\frac{1}{2}\right), \label{basis}
\ee
where $\{\ket{\psi_n}$ is a state vector of harmonic oscillator with well known coordinate wave function:
\be
\psi_n (x) = \frac{1}{\sqrt{2^n n!}}\left(\frac{m\omega}{\pi}\right)^\frac{1}{4}e^{-\frac{m\omega x^2}{2}} \mathbb{H}_n \left(\sqrt{m\omega} x\right). \label{oscherm}
\ee
$\mathbb{H}_n (x)$ are Hermite polynomials, and as usual, $\hbar$ was taken to be equal to 1.
In turn, spin eigenvectors and eigenvalues of $\sigma_3$ are as follows:
\be
\ket{0} = \begin{pmatrix}
1 \\
0
\end{pmatrix};\quad \ket{1} = \begin{pmatrix}
0 \\
1
\end{pmatrix};\quad \sigma_3 \ket{a} = (-1)^a \ket{a}, \:\:a=0,1.
\ee

For brevity, hereinafter we will use the unique index $M \equiv (m,a),$ where $m$ is from the oscillator space and $a$ is from the spin space. Correspondingly, eigenvectors of $H$ are now $\{\ket{\Psi_M}\} = \{\ket{\psi_n}\otimes \ket{a}\}.$ Two indices coincide $M=M'$ only if $m=m',\, a=a'.$
%Inequality $m \neq n$ is being replaced by $(m,k)\neq(n,l)$. It should be understood as the whole pair is not equal to the other one. For example, it could be $m\neq n$ but %$k=l$ and it would still be the case. And $(m,k) = (n,l)$ means that $m=n$ and $k=l$.

Accordingly, an arbitrary operator $O$ can be represented by the coefficients $o_{MN}$ as
\ba
O &=& \sum_{M,N} o_{MN}\ket{\Psi_M}\bra{\Psi_N} = \sum o_{MN} \ket{\psi_m}\bra{\psi_n} \otimes \ket{a} \bra{b}; \label{product}\\
M&=&(m, a),\, N=(n, b).  \nonumber
\ea

From the preceding relations it follows that
\be
H\ket{\Psi_M} = \left(\omega\left(m+\frac{1}{2}\right)+\delta(-1)^a\right)\ket{\Psi_M}, \nonumber
\ee
and the corresponding matrix elements of $H$ are
\be
\eps_{MN} = \eps_M\delta_{MN} = \delta_{mn}\delta_{ab}\left(\omega\left(m+\frac{1}{2}\right)+\delta(-1)^a\right). \label{epsmk}
\ee
The matrix elements of $L$ are as follows:
\be
l_{MN}\equiv l_{mnab} = \lambda\delta_{mn}\delta_{a0}\delta_{b1}. \label{lmnkl}
 \ee

Let us write \textbf{Equations 1} (\ref{main1}) and \textbf{2} (\ref{main2}) for this system, keeping in mind the general scheme of calculation for degenerate Hamiltonian described above:\\
\textbf{Equation 1}:
\ba
&&\sum_{P} |l_{MP}|^2 f^{(s)}_{PP} - \sum_{P} |l_{PM}|^2 f^{(s)}_{MM} +
 \sum_{PR;P\neq R}l_{MP}f^{(s)}_{PR}l^{*}_{MR} \nonumber\\
&&- \frac{1}{2}\sum_{PR;R\neq M}l^{*}_{PM}l_{PR}f^{(s)}_{RM}- \frac{1}{2}\sum_{PR;R\neq M}f^{(s)}_{MR}l^{*}_{PR}l_{PM} = 0  \label{ex1eq1pert}
\ea
for arbitrary fixed index $M=(m,a)$.\\
\textbf{Equation 2}:
\ba
&&-i f^{(s)}_{MN}(\eps_{M}-\eps_{N}) + \lambda^2\sum_{P} l_{MP}f^{(s-1)}_{PP} l^{*}_{NP}  \nonumber\\
&& - \frac{1}{2}\lambda^2\sum_{P} l^{*}_{PM}l_{PN}(f^{(s-1)}_{MM}+f^{(s-1)}_{NN}) + \lambda^2\sum_{PR;P\neq R}l_{MP}f^{(s-1)}_{PR}l^{*}_{NR} \nonumber\\
&&- \frac{1}{2}\lambda^2\sum_{PR;P\neq R}l^{*}_{PM}l_{PR}f^{(s-1)}_{RN}- \frac{1}{2}\lambda^2\sum_{PR;R\neq M}f^{(s-1)}_{MR}l^{*}_{PR}l_{PN} = 0 \label{ex1eq2pert}
\ea
for arbitrary fixed pairs $M=(m,a)$ and $N=(n,b)$, $M\neq N$.

Substituting $\eps_{M}$ (\ref{epsmk}) and $l_{MN}$ (\ref{lmnkl}) and simplifying the equations, we get\\
\textbf{Equation 1}:
\be
f^{(s)}_{MM}=0  \label{easyosc1}
\ee
for $M=(m,1)$.
\\
and \textbf{Equation 2}:
\ba
&&-i f^{(s)}_{MN} (\eps_{M} - \eps_{N}) + |\lambda|^2 \Big( f^{(s-1)}_{mn11}\delta_{a0}\delta_{b0}(1-\delta_{mn})  \nonumber\\
&&- \frac{1}{2} f^{(s-1)}_{mn1b}\delta_{a1}(1-\delta_{mn}\delta_{1b}) - \frac{1}{2} f^{(s-1)}_{mna1}\delta_{b1}(1-\delta_{mn}\delta_{a1}) \Big) = 0. \label{easyosc2}
\ea
for all pairs $M=(m,a)$ and $N=(n,b)$, such that $M\neq N$.

To separate non-degenerate and degenerate cases in this model, let us find the pairs of indices $M=(m,a)$ and $N=(n,b)$ which are "internal" according to the terminology introduced above, i.e., $\eps_M=\eps_N.$
Recalling (\ref{epsmk}),
one can conclude that the only options to get degeneracy are:
\ba
&&M=(m,0)\:\:\:\text{and}\:\:\:N=(m+q,1); \quad q\equiv\frac{2\delta}{\omega} \label{int1} \\
&&M=(m,1) \:\:\:\text{and}\:\:\: N=(m-q,0); \quad m=0,1,2,\dots . \label{int2}
\ea
which are possible only if $q=\frac{2\delta}{\omega}$ is an integer.

\subsubsection{Non-degenerate Hamiltonian}\label{sec411}
Let the parameter $q$ is not an integer.
The form of \textbf{Equation 1} looks just as (\ref{easyosc1}),
and, according to (\ref{easyosc2}),
\textbf{Equation 2} is,
\ba
&&f^{(s)}_{MN} = -\frac{i|\lambda|^2}{\eps_{M} - \eps_{N}}  \Big( f^{(s-1)}_{mn11}\delta_{a0}\delta_{b0}(1-\delta_{mn})  \nonumber\\
&&- \frac{1}{2} f^{(s-1)}_{mn1b}\delta_{a1}
(1-\delta_{mn}\delta_{b1}) - \frac{1}{2} f^{(s-1)}_{mna1}\delta_{b1}(1-\delta_{mn}\delta_{a1}) \Big) \label{easyosc22}
\ea
for the pairs $M=(m,a)$ and $N=(n,b)$, such that $M\neq N.$

Let us follow the scheme of calculation for non-degenerate Hamiltonian described earlier.
\begin{itemize}
\item We look at \textbf{Equation 2} (\ref{easyosc22}) and see that $f^{(0)}_{MN}=0,\, M\neq N$, since in the right part $f^{(-1)}_{PR}=0$ for any indices $P, R.$
\item \textbf{Equation 1} (\ref{easyosc1}) states that $f^{(0)}_{mm11}=0$ and, consequently, $f^{(0)}_{mm00}$ are arbitrary real (because of hermiticity of the density matrix) numbers  only restricted by the trace condition $\sum_m f^{(0)}_{mm00} =1$.
\item We must substitute all matrix elements of the zeroth order - diagonal elements $f^{(0)}_{mm00}$, $f^{(0)}_{mm11}$ and non-diagonal elements $f^{(0)}_{mnab}, {M\neq N}$ - into  \textbf{Equation 2} (\ref{easyosc22}) in order to find first-order non-diagonal elements $f^{(1)}_{MN}, {M\neq N}$. The only elements of the zeroth order that do not vanish are $f^{(0)}_{mm00}$.
    %In the right part of equation 2 (\ref{easyosc22}) are only $f_{mnkl}$ with one or two of the last two indexes equal to $1$.
    Thus, the right part of \textbf{Equation 2} (\ref{easyosc22}) is equal to $0$, i.e. $f^{(1)}_{MN} = 0, {M\neq N}.$
\item Keeping in mind (\ref{easyosc1}), we know that $f^{(1)}_{mm11}=0$, therefore $f^{(1)}_{mm00}$ are arbitrary real numbers only restricted by the trace condition\\
{$\sum_m f^{(1)}_{mm00} =0.$}
\item Again the right part of  \textbf{Equation 2} (\ref{easyosc22}) is equal $0$. Therefore,\\
$f^{(2)}_{MN} = 0, {M\neq N}.$
\item And so on.
\end{itemize}

Summing up,
$f_{mm00}$ are arbitrary real numbers only restricted by the trace condition $\sum_m f_{mm00} =1$, while all other matrix elements $f_{MN}$ vanish.

If we recall the {\it Liouville} pointers, they correspond to diagonal density matrices with arbitrary elements, that means, with arbitrary $f_{mm00}$ and $f_{mm11}$ only obeying the trace condition: $\sum_m (f_{mm00} + f_{mm11}) = 1$. We demonstrated above that the {\it Lindblad} pointers differ essentially from the {\it Liouville} ones: while the part $f_{mm00}$ of diagonal elements of the density matrix after turning on an interaction with an environment may coincide with those without interaction, the part $f_{mm11}$ definitely vanishes. We can say that after turning an interaction on, the resulting density matrix "receives the direction", namely, along $f_{mm11}=0.$ If vice versa, we decide to turn off the interaction with an environment, the resulting density matrix "remembers" it keeping $f_{mm11}=0.$

\subsubsection{Degenerate Hamiltonian}

In this case the incoming parameters are such that $\frac{2\delta}{\omega}$ is an integer.
According to the scheme of calculation for degenerate Hamiltonian, we need two versions of \textbf{Equation 2} (\ref{easyosc2}): one - for internal ((\ref{int1}), (\ref{int2})) pairs of indices, and second - for external pairs. The result is:\\
\textbf{Equation 1}:
\be
f^{(s)}_{mm11}=0 \label{easyoscdeg1}
\ee
%for some pair $(m,k)$.
\\
\textbf{Equation 2}:
\begin{itemize}
\item
For internal pairs:
\be
f^{(s)}_{m,m+q,0,1}=0; \quad f^{(s)}_{m,m-q,1,0}=0. \label{easyoscdeg21}
\ee

\item
For external pairs $M=(m,a),\, N=(n,b),\, M\neq N$:
\ba
&&f^{(s)}_{MN} = -\frac{i|\lambda|^2}{\eps_{M} - \eps_{N}} \Big( f^{(s-1)}_{mn11}\delta_{a0}\delta_{b0}(1-\delta_{mn}) \nonumber\\
&&-\frac{1}{2} f^{(s-1)}_{mn1b}\delta_{a1}(1-\delta_{mn}\delta_{1b}) - \frac{1}{2} f^{(s-1)}_{mna1}\delta_{b1}(1-\delta_{mn}\delta_{a1}) \Big). \label{easyoscdeg23}
\ea
%for $M=(m,a),\, N=(n,b),\, M\neq N.$
\end{itemize}

Let us find all the orders $f^{(s)}_{mnab}$ now, following the procedure above.
\begin{itemize}
\item First we look at \textbf{Equation 2} for external pairs (\ref{easyoscdeg23}) and obtain that $f^{(0)}_{MN}=0$ for external $M\neq N,$ since $f^{(-1)}_{PR}=0$ for all possible indices $P, R.$
\item Then, in \textbf{Equation 1} (\ref{easyoscdeg1}) and \textbf{Equation 2} for internal indices (\ref{easyoscdeg21}), we insert $f^{(0)}_{MN}=0$ for external $M\neq N$ in order to find diagonal $f^{(0)}_{MM}$ and the rest of non-diagonal $f^{(0)}_{MN}=0$ for internal $M\neq N.$ It is evident from (\ref{easyoscdeg1}) and (\ref{easyoscdeg21}) that:\\
1. Diagonal elements:  $f^{(0)}_{mm11}=0$;  $f^{(0)}_{mm00}$ are arbitrary except for the only limitation: $\sum_m f^{(0)}_{mm00}=1$ (trace condition).\\
2. Non-diagonal internal elements: $f^{(0)}_{m,m+q,0,1}=0$, $f^{(0)}_{m,m-q,1,0}=0$ (there are no other internal pairs of indices).
\item We look again at \textbf{Equation 2} for external pairs of indices (\ref{easyoscdeg23}). We need to insert there all matrix elements of the zeroth order $f^{(0)}_{MN}$. But we know that the only nonvanishing ones are $f^{(0)}_{mm00}.$ Therefore, the r.h.s. is equal to $0$, and $f^{(1)}_{MN}=0$ for external $M\neq N.$
\item We analyze \textbf{Equations 1} (\ref{easyoscdeg1}) and \textbf{2} for internal pairs of indices (\ref{easyoscdeg21}). We again easily find:\\
1. Diagonal elements:  $f^{(1)}_{mm11}=0$;  $f^{(1)}_{mm00}$ are arbitrary except for the limitation: $\sum_m f^{(1)}_{mm00}=0$ (trace condition).\\
2. Non-diagonal internal elements: $f^{(1)}_{m,m+q,0,1}=0$, $f^{(1)}_{m,m-q,1,0}=0$. There are no other internal pairs of indices.
\item Having a look at \textbf{Equation 2} for external pairs of indices (\ref{easyoscdeg23}), again we conclude that $f^{(2)}_{MN}=0$ for external $M\neq N.$
\item And so on.
%etc.
\end{itemize}

As a result, we have that $f_{mm00}$ are arbitrary except for the trace condition: $\sum_m f_{mm00}=1.$ All the other matrix elements vanish.

Let us again compare the {\it Liouville} and the {\it Lindblad} pointers. The first ones are the density matrices with arbitrary diagonal elements $f_{mm00}$ and $f_{mm11}$, restricted only by the trace condition $\sum_m (f_{mm00} + f_{mm11}) = 1$, and with arbitrary elements of the kind $f^{(s)}_{m,m+q,0,1}$ and $f^{(s)}_{m,m-q,1,0}$ (indices $(m,0)$ and $(m+q,1)$ correspond to the states with the same energy, the same concerns $(m,1)$ and $(m-q,0)$). We see that interaction with an environment leads to the {\it Liouville} pointer with destroyed degeneracy, in addition to disposing of $f_{mm11}$, as in the non-degenerate case (see Subsect. \ref{sec411}).

\subsection{Second oscillator example}
As a second example, we shall take the same system of one-dimensional harmonic oscillator with spin $1/2$ in a magnetic field presented by the Hamiltonian (\ref{H}) but with another Lindblad operator. Instead of a single operator $L$ along $\sigma_+=\sigma_1+i\sigma_2,$ we shall use two Lindblad operators $L^{(1)}, L^{(2)}$ directed along $\sigma_1, \sigma_2,$ correspondingly:
\be
L^{(1)} = \gamma_1\cdot I \otimes \sigma_1;\quad L^{(2)} = \gamma_2\cdot I \otimes \sigma_2, \label{l1}
\ee
where $\gamma_1$ and $\gamma_2$ are arbitrary (in general, complex) small numbers \\ $|\gamma_{1,2}| \ll 1.$
%the system cannot diverge). Even if $\gamma_1$ and $\gamma_2$ are complex numbers, the proof still works. So, our goal is to find all possible pointers, final stationary states.
We shall keep the same definitions and notations as in Example 1, i.e., Eqs.(\ref{basis}) - (\ref{epsmk}) unchanged, but with the following expressions for the matrix elements
of the Lindblad operators:
\ba
L^{(1)}_{MN}&=&l^{(1)}_{mnab}= \gamma_1 \delta_{mn}(\delta_{a0}\delta_{b1}+ \delta_{a1}\delta_{b0});  \nonumber\\
L^{(2)}_{MN}&=&l^{(2)}_{mnab}=\gamma_2 \delta_{mn}(-i \delta_{a0}\delta_{b1}+ i \delta_{a1}\delta_{b0}).     \label{lmnab1}
%\ee
%\be \label{lmnab2}
%L^{(2)}_{MN}=\gamma_2 \delta_{mn}(-i \delta_{a0}\delta_{b1}+ i \delta_{a1}\delta_{b0})
\ea
Again, the spectrum of the Hamiltonian (\ref{H}) can be degenerate for integer values of $q=\frac{2\delta}{\omega}.$ In such a case, due to (\ref{epsmk}) the "internal" pairs of degenerate levels
%with eigenvalues
%\be
%\be \omega (m-n) + \delta [(-1)^k - (-1)^l] = 0 \ee
%\ee
are again as in (\ref{int1}), (\ref{int2}).
%
%Let us find the pairs of indexes $(m,k)$ and $(n,l)$ which are ''internal'' according to introduced terminology in order to subsequently apply our calculation scheme for a %degenerate Hamiltonian.
%
%The pair $(m,k)$ is internal with the pair $(n,l)$, if $\eps_{mk} = \eps_{nl}$, that is (recalling (\ref{epsmk}))
%\be \omega (m-n) + \delta [(-1)^k - (-1)^l] = 0 \ee
%
%Looking over all possible combinations of values of $k$ and $l$ ($k,l=0,1$), we see that the only unequal pairs $(m,k)$ and $(n,l)$ that satisfy this equation are those:
%\be \label{int1} (m,0)\:\:\:\text{and}\:\:\:(m+\frac{2\delta}{\omega},1) \ee
%\be \label{int2} (m,1) \:\:\:\text{and}\:\:\: (m-\frac{2\delta}{\omega},0) \ee
%
%Here $m$ is any possible number, corresponding to the oscillator level.
%
%We see that these pairs are meaningful only if $\frac{2\delta}{\omega}$ is an integer. Therefore, our problem divides into 2 subproblems:\\
%
%1. $\frac{2\delta}{\omega}$ is not an integer. Hamiltonian is non-degenerate.\\
%\begin{spacing}{0.2}
%2. $\frac{2\delta}{\omega}$ is an integer. Hamiltonian is degenerate.
%\end{spacing}

%\subsubsection{Lindblad and Liouville pointers}
Let us now consider the situation with degenerate Hamiltonian postponing non-degenerate case up to the end of the subsection.
%We will come back to the first option later.
Substituting $\eps_{M}$ (\ref{epsmk}) and $L^{(1),(2)}_{MN}$ (\ref{lmnab1})
%and $l^{(2)}_{mnab2}$ (\ref{lmnab2})
into \textbf{Equations 1} and \textbf{2} (Eqs. (\ref{deg1}) and (\ref{deg21})-(\ref{deg22}), correspondingly), simplifying them, and including an additional summation over the set of the Lindblad operators $L^{(a)},\, a=1,2,$ we get
\textbf{(here and below, matrix elements with internal pairs of indices are in bold)}
\\
\textbf{Equation 1}:
\be
\label{eq1} f^{(s)}_{mm00}=f^{(s)}_{mm11}
\ee
\textbf{Equation 2}:
\begin{itemize}
\item For internal pairs of indices:
\be \label{eq21}\boldsymbol{f^{(s)}_{m,m+q,0,1}}=\frac{|\gamma_1|^2-|\gamma_2|^2}{|\gamma_1|^2+|\gamma_2|^2} f^{(s)}_{m,m+q,1,0}\ee
\be \label{eq22}\boldsymbol{f^{(s)}_{m,m-q,1,0}}=\frac{|\gamma_1|^2-|\gamma_2|^2}{|\gamma_1|^2+|\gamma_2|^2} f^{(s)}_{m,m-q,0,1}\ee

\item And for external pair of indices:
\be
\label{eq23} f^{(s)}_{m,m-q,0,1} = -\frac{i}{4\delta} \Big[ \left(|\gamma_1|^2 - |\gamma_2|^2\right) \boldsymbol{f^{(s-1)}_{m,m-q,1,0}} - \left(|\gamma_1|^2 + |\gamma_2|^2\right)f^{(s-1)}_{m,m-q,0,1} \Big];
\ee
\be
\label{eq24} f^{(s)}_{m,m+q,1,0} = \frac{i}{4\delta} \Big[ \left(|\gamma_1|^2 - |\gamma_2|^2\right) \boldsymbol{f^{(s-1)}_{m,m+q,0,1}} - \left(|\gamma_1|^2 + |\gamma_2|^2\right)f^{(s-1)}_{m,m+q,1,0} \Big];
\ee
\ba
f^{(s)}_{mn01} &=& -\frac{i}{\omega(m-n)+2\delta} \Big[ \left(|\gamma_1|^2 - |\gamma_2|^2\right) f^{(s-1)}_{mn10} - \left(|\gamma_1|^2 + |\gamma_2|^2\right)f^{(s-1)}_{mn01} \Big], \nonumber\\
n &\neq & m\pm q; \label{eq25}
\ea
\ba
f^{(s)}_{mn10} &=& -\frac{i}{\omega(m-n)-2\delta} \Big[ \left(|\gamma_1|^2 - |\gamma_2|^2\right) f^{(s-1)}_{mn01} - \left(|\gamma_1|^2 + |\gamma_2|^2\right)f^{(s-1)}_{mn10} \Big], \nonumber\\
 n& \neq & m\pm q; \label{eq26}
\ea
\be
\label{eq27} f^{(s)}_{mn00} = -\frac{i\left( |\gamma_1|^2 + |\gamma_2|^2 \right)}{\omega(m-n)} \Big[f^{(s-1)}_{mn11} - f^{(s-1)}_{mn00}\Big], m \neq n;
\ee
\be
\label{eq28} f^{(s)}_{mn11} = \frac{i\left( |\gamma_1|^2 + |\gamma_2|^2 \right)}{\omega(m-n)} \Big[f^{(s-1)}_{mn11} - f^{(s-1)}_{mn00}\Big], m \neq n.
\ee
%for external pairs of indices.
\end{itemize}
%\textbf{Matrix elements with internal pairs of indices} are in bold above.

In this example, we shall demonstrate how our general scheme of calculation for degenerate Hamiltonian works.
All $f^{(s)}_{mnab}$ with external pairs of indices $(m,a)$ and $(n,b)$ are found from Eqs. (\ref{eq23})-(\ref{eq28}), while $f^{(s)}_{mnab}$ with internal pairs of indices, that is $\boldsymbol{f^{(s)}_{m,m+q,0,1}}$ and $\boldsymbol{f^{(s)}_{m,m-q,1,0}}$, are found from Eqs. (\ref{eq21}), (\ref{eq22}). Diagonal elements $f^{(s)}_{mmaa}$ are found from (\ref{eq1}). Therefore,

\begin{itemize}
\item
First, we find from Eqs. (\ref{eq23})-(\ref{eq28}) that $f^{(0)}_{mnab} = 0$, if $(m,a)$, $(n,b)$ are external pairs ({\it i.e.,}
%all that we have in the right parts of these equations is
%f^{(-1)}_{mnab}$, and $f^{(-1)}_{mnab}=0$
for all possible values of indices).
\item
Then, we must put $f^{(0)}_{mnab} = 0$ for external pairs $(m,a)$ and $(n,b)$
%are external pairs$\}$
into Eqs. (\ref{eq21}), (\ref{eq22}), so we find that $\boldsymbol{f^{(0)}_{m,m+q,0,1}}$ and $\boldsymbol{f^{(0)}_{m,m-q,1,0}}$ also vanish.
\item
$f^{(0)}_{MN}=0$, for $M\neq N$.
%if $(m,a)$ and $(n,b)$ are non-equal pairs.
Only this kind of matrix elements is in the right parts of (\ref{eq23})-(\ref{eq28}) (if we consider these equations in the next order of perturbation theory). Therefore, $f^{(1)}_{MN} = 0$, if $(m,a)$, $(n,b)$ are external pairs.
\item
Analogously, from Eqs. (\ref{eq21}), (\ref{eq22}) it follows that $\boldsymbol{f^{(1)}_{m,m+q,0,1}}$ and \\ $\boldsymbol{f^{(1)}_{m,m-q,1,0}}$ vanish.
\item
Then, we need to put $f^{(1)}_{MN}=0$ for $M\neq N$ in the r.h.s. of (\ref{eq23})-(\ref{eq28}), and so on.\\
\end{itemize}
%Finally, we conclude that \underline{$f_{MN}=0$ for non-equal pairs $M\neq N$,} since $f^{(s)}_{MN}=0$ for each order $s$ of perturbation theory.
Finally, we conclude that $f_{MN}=0$ for non-equal pairs $M\neq N$, since $f^{(s)}_{MN}=0$ for each order $s$ of perturbation theory.
At last, we need to deal with $f_{MM}$. Eq. (\ref{eq1}) simply gives that $f_{mm00}=f_{mm11}$, since $f^{(s)}_{mm00}=f^{(s)}_{mm11}$ in each order $s$ of perturbation theory. It means that all matrix elements $f_{mm00}$ are arbitrary, $\{f_{mm11}\}$ are defined from them. Next, the arbitrariness is reduced by the trace condition: $\sum_m f_{mm00}=\frac{1}{2}$ (it follows from $\sum_{ma} f_{mmaa}=1$).
%At last, we need to deal with $f_{MM}$. Equation (\ref{eq1}) simply gives that \underline{$f_{mm00}=f_{mm11}$}, since $f^{(s)}_{mm00}=f^{(s)}_{mm11}$ in each order $s$ of %perturbation theory. It means that \underline{all matrix elements $f_{mm00}$ are arbitrary}, $\{f_{mm11}\}$ are defined from them. Next, the arbitrariness is reduced by the %trace condition: \underline{$\sum_m f_{mm00}=\frac{1}{2}$} (it follows from $\sum_{ma} f_{mmaa}=1$).

If we consider now the non-degenerate case, we will get the same results. The only difference
%What changes
is that in this case matrix elements $\boldsymbol{f^{(s)}_{m,m+q,0,1}}$, $f^{(s)}_{m,m+q,1,0}$ and so on must be skipped (together with Eqs. (\ref{eq21})-(\ref{eq24})), since they do not exist for non-integer values of $q=\frac{2\delta}{\omega}$. Nonetheless, the reasoning stays the same, so the result %abide.
still is true.

%\subsubsection{Liouville pointers}
As we have argued in the very end of Sect. \ref{sec2} for the general case,
%derived before,
the Liouville pointers are density matrices with arbitrary diagonal elements and elements with unequal internal indices. Thus, in the example considered here Liouville pointers are density matrices with arbitrary $f_{mm00}$, $f_{mm11}$ and $\boldsymbol{f_{m,m+q,0,1}}$, $\boldsymbol{f_{m,m-q,1,0}}$ (if Hamiltonian is degenerate). ${\{f_{mm00}, f_{mm11}\}}$ are restricted by the trace condition: ${\sum_m f_{mm00} + \sum_m f_{mm11} = 1}$.

%\subsubsection{Comparison between Lindblad and Liouville pointers}
It is interesting to compare the forms of {\it Lindblad} and {\it Liouville} pointers for the considered system. This will be done with the help of two tables: separately for degenerate and non-degenerate Hamiltonians.
\begin{itemize}
\item Degenerate Hamiltonian ($q=\frac{2\delta}{\omega}$ is an integer)
\begin{center}
 \begin{tabular}{|l|l|}
 \hline
 Liouville pointers & Lindblad pointers \\
 \hline
$f_{mm00}$ are arbitrary*,  &  $f_{mm00}$ are arbitrary*, \\
$f_{mm11}$ are arbitrary*,  &  $f_{mm11}=f_{mm00}$  \\
$\boldsymbol{f_{m,m+q,0,1}}$ are arbitrary, & \\
$\boldsymbol{f_{m,m-q,1,0}}$ are arbitrary & \\
*but restricted by the trace condition   &   *but restricted by the trace condition \\
$\sum_m f_{mm00} + \sum_m f_{mm11} =1 $  &  $\sum_m f_{mm00} = \frac{1}{2}$ \\
\hline
\end{tabular}
\end{center}

We know that when we turn on an interaction of a system with environment (described in this example by Lindblad operators (\ref{l1})), Liouville pointers (expected final quantum states of a closed system) transform into Lindblad pointers (expected final quantum states of an open system). Thus, our result shows that this kind of interaction aligns the populations of the spin up and spin down quantum states. Moreover, this interaction completely destroys degeneracy.

\item Non-degenerate Hamiltonian ($q=\frac{2\delta}{\omega}$ is \textbf{not} an integer)
\begin{center}
 \begin{tabular}{|l|l|}
 \hline
 Liouville pointers & Lindblad pointers \\
 \hline
$f_{mm00}$ are arbitrary*,  &  $f_{mm00}$ are arbitrary*, \\
$f_{mm11}$ are arbitrary*,  &  $f_{mm11}=f_{mm00}$  \\
*but restricted by the trace condition   &   *but restricted by the trace condition \\
$\sum_m f_{mm00} + \sum_m f_{mm11} =1 $  &  $\sum_m f_{mm00} = \frac{1}{2}$ \\
\hline
\end{tabular}
\end{center}

The conclusion is the same. Populations of spin up and spin down states become equal as a result of the interaction of the system with the environment.
\end{itemize}
What is interesting, the result does not depend on the values of coupling constants $\gamma_1,\,\gamma_2$ (see (\ref{l1})).
%It is only important that at least one of the operators $L_1$ or $L_2$ is not equal to $0$.

\subsection{Third example: two-dimensional space of states}
Let us consider an example, where a Hamiltonian and a single Lindblad operator are given by the following $2\times 2$ matrices in the energetic basis like in (\ref{h}) - (\ref{rho0_0}).
$H$ is non-degenerate:
\be
H = \begin{pmatrix}
\eps_1 & 0 \\
0 & \eps_2
\end{pmatrix}, \eps_1 \neq \eps_2, \label{HH}
\ee
and $L$ - off-diagonal matrix:
%that the proposed scheme of calculations gives reasonable results for pointers.\\
%We take $L$ as $2\times 2$ off-diagonal matrix:
\be
L = \begin{pmatrix}
0 & l_{12} \\
l_{21} & 0
\end{pmatrix} . \label{LL}
\ee

\subsubsection{
Pointers for FGKLS equation by means of perturbation theory}

\begin{enumerate}
\item
The general expression (\ref{nondeg2}) of Sect. \ref{sec3} for non-diagonal elements
%\be f^{(s)}_{mn} = \frac{i}{\eps_m-\eps_n}\Big[ -\sum_a\sum_k l^{[a]}_{mk}f^{(s-1)}_{kk} l^{[a]*}_{nk} + \frac{1}{2}\sum_a\sum_k %l^{[a]*}_{km}l^{[a]}_{kn}(f^{(s-1)}_{mm}+f^{(s-1)}_{nn}) - \nonumber\ee
%\be \label{nondeg2}- \sum_a\sum_{k,l;k\neq l}l^{[a]}_{mk}f^{(s-1)}_{kl}l^{[a]*}_{nl} + \frac{1}{2}\sum_a\sum_{k,l;l\neq %n}l^{[a]*}_{km}l^{[a]}_{kl}f^{(s-1)}_{ln}+\frac{1}{2}\sum_a\sum_{k,l;l\neq m}f^{(s-1)}_{ml}l^{[a]*}_{kl}l^{[a]}_{kn}\Big], m \neq n\ee
is essentially simplified now
%\begin{itemize}
since first two sums in the r.h.s. vanish automatically for arbitrary off-diagonal two-dimensional matrix $l_{kl}.$
%does not equal $0$, only if $k\neq l$. Therefore, in the first two terms in the right part of the equation it should be $k\neq m$ and $k\neq n$. Since the dimension is $2$, %it follows that $m=n$ which cannot be true. Thus, these terms disappear.\\
Finally, this equation can be written as:
\be
f^{(s)}_{12} = \frac{i}{\eps_1 - \eps_2} \left[ -l_{12} f^{(s-1)}_{21}l^*_{21} + \frac{1}{2} (|l_{21}|^2 +  |l_{12}|^2)  f^{(s-1)}_{12}\right]. \label{fff}
\ee
The equation for $f^{(s)}_{21}$ is similar due to hermiticity of $\rho .$
%: it can easily be retrieved, considering $f^{(s)}_{21} = f^{(s)*}_{12}$.\\
Since \\$f^{(-1)}_{12}=f^{(-1)}_{21}=0$, we obtain from (\ref{fff}) that $f^{(0)}_{12}= f^{(0)}_{21}=0,$
%, if we just substitute the former in the right part of the equation for $s=0$.
and all the orders $f^{(s)}_{12}=f^{(s)}_{21} = 0,$ as well. Thus, all non-diagonal elements of $\rho$ vanish.
%\end{itemize}

\item
The Eq. (\ref{nondeg1}) for diagonal elements of $\rho$ becomes also very simple in this example due to vanishing of off-diagonal elements $f_{ij},\, i\neq j$ so that
%\be
%\begin{pmatrix}
%-\sum_{t\neq 1}|l_{t1}|^2 & |l_{12}|^2 & |l_{13}|^2 & \dots \\
%|l_{21}|^2 & -\sum_{t\neq 2}|l_{t2}|^2 & |l_{23}|^2 & \dots\\
%|l_{31}|^2 & |l_{32}|^2 & -\sum_{t\neq 3}|l_{t3}|^2 & \dots \\
%\vdots & \vdots & \vdots & \ddots \\
%\end{pmatrix}
%\begin{pmatrix}
%f^{(s)}_{11} \\
%f^{(s)}_{22} \\
%f^{(s)}_{33} \\
%\vdots \\
%\end{pmatrix}
%= \nonumber
%\ee
%\be \label{nondeg1}=
%\begin{pmatrix}
%-\sum_{k,l;k\neq l}l_{1k}f^{(s)}_{kl}l^*_{1l} + \frac{1}{2}\sum_{k,l;l\neq 1}l^*_{k1}l_{kl}f^{(s)}_{l1}+\frac{1}{2}\sum_{k,l;l\neq 1}f^{(s)}_{1l}l^*_{kl}l_{k1} \\
%-\sum_{k,l;k\neq l}l_{2k}f^{(s)}_{kl}l^*_{2l} + \frac{1}{2}\sum_{k,l;l\neq 2}l^*_{k2}l_{kl}f^{(s)}_{l2}+\frac{1}{2}\sum_{k,l;l\neq 2}f^{(s)}_{2l}l^*_{kl}l_{k2}  \\
%-\sum_{k,l;k\neq l}l_{3k}f^{(s)}_{kl}l^*_{3l} + \frac{1}{2}\sum_{k,l;l\neq 3}l^*_{k3}l_{kl}f^{(s)}_{l3}+\frac{1}{2}\sum_{k,l;l\neq 3}f^{(s)}_{3l}l^*_{kl}l_{k3}  \\
%\vdots \\
%\end{pmatrix}
%\ee
%Given the latter result, we see that the right part of the equation vanishes. Finally, it can be simplified to
the system becomes homogeneous:
\be \begin{pmatrix}
-|l_{21}|^2 & |l_{12}|^2 \\
|l_{21}|^2 &  -|l_{12}|^2
\end{pmatrix}
\begin{pmatrix}
f^{(s)}_{11}\\
f^{(s)}_{22}
\end{pmatrix} = 0\ee
\end{enumerate}

Applying the trace conditions (\ref{trace+}), we get expressions for diagonal elements of $\rho :$
\be
f_{11} = \frac{|l_{12}|^2}{|l_{12}|^2+|l_{21}|^2};
\ee
\be
f_{22} = \frac{|l_{21}|^2}{|l_{12}|^2+|l_{21}|^2}.
\ee

Summing up, the only
%possible pointer,
solution for the pointer of FGKLS equation in our scheme of calculation is:
\be
\label{pertpo}\begin{pmatrix}
\frac{|l_{12}|^2}{|l_{12}|^2+|l_{21}|^2} & 0 \\
0 & \frac{|l_{21}|^2}{|l_{12}|^2+|l_{21}|^2}
\end{pmatrix}.
\ee

\subsubsection{
Pointers by means of exact solution}
An arbitrary diagonal non-degenerate Hamiltonian (\ref{HH}) and off-diagonal $L$ as in (\ref{LL}) can be expressed as:
\ba
H &=& \eps_0 I + \eps_3 \sigma_3, \;\;\; \eps_0 \neq \eps_3 \in \mathbb{R}; \label{HHH}\\
%For it to be non-degenerate, it should be $\eps_3 \neq 0$.\\
%An arbitrary off-diagonal $L$ as in (\ref{LL} can be expressed as
L &=& (a_1 + i b_1) \sigma_1 + (a_2 + i b_2) \sigma_2, \;\;\; a_1, a_2, b_1, b_2 \in \mathbb{R}.\label{LLL}
\ea
Correspondingly, the FGKLS equation (\ref{lindblad}) takes the form:
%obtained equation
\be
\label{eq} \dot{\pmb{\rho}} = 2 \left[\pmb{h} - a_0 \pmb{b} + b_0  \pmb{a}, \pmb{\rho} \right] - 2| \pmb{a}|^2  \pmb{\rho_{\perp a}} - 2| \pmb{b}|^2 \pmb{\rho_{\perp b}} + 2 \left[ \pmb{a}, \pmb{b} \right]
\ee
where $3$-vector $\pmb{\rho}=\left(\rho_1,\rho_2,\rho_3\right)$,
\be
 \pmb{\rho_{\perp a}} = \pmb{\rho} - \frac{( \pmb{a} \cdot \pmb{\rho})}{| \pmb{a}|^2} \pmb{a}
\ee
and in components (\ref{eq}) reads:
\ba
&&\dot{\rho_1} = -2(a_2^2 + b_2^2) \rho_1 + 2  (-\eps_3 + a_1 a_2 + b_1 b_2) \rho_2; \label{11}\\
&&\dot{\rho_2} = 2 (\eps_3 + a_1 a_2 + b_1 b_2)\rho_1 -2(a_1^2 + b_1^2) \rho_2;  \label{22}\\
&&\dot{\rho_3} = -2(a_1^2 + a_2^2 + b_1^2 + b_2^2)\rho_3 + 2(a_1 b_2 - a_2 b_1). \label{33}
\ea

\begin{itemize}
\item
The last Eq. (\ref{33}) can be solved straightforwardly:
%the general solution is the sum of the particular solution of the non-homogeneous equation plus the general solution of the homogeneous equation:
\be
\rho_3(t) = \frac{a_1 b_2 - a_2 b_1}{a_1^2 + a_2^2 + b_1^2 + b_2^2} + \left(\rho_3(0) - \frac{a_1 b_2 - a_2 b_1}{a_1^2 + a_2^2 + b_1^2 + b_2^2}\right) e^{-2(a_1^2 + a_2^2 + b_1^2 + b_2^2)t}
\ee
%So, the limit value is
with the limit value:
\be
\rho_3(+\infty) = \frac{a_1 b_2 - a_2 b_1}{a_1^2 + a_2^2 + b_1^2 + b_2^2}. \label{aaa}
\ee
which provides the diagonal part of the pointer $\frac{1}{2}I + \rho_3(+\infty) \sigma_3.$
%\be \rho(+ \infty) =  \begin{pmatrix}
%\frac{1}{2} + \frac{a_1 b_2 - a_2 b_1}{a_1^2 + a_2^2 + b_1^2 + b_2^2} & \text{?} \\
%\text{?} & \frac{1}{2} - \frac{a_1 b_2 - a_2 b_1}{a_1^2 + a_2^2 + b_1^2 + b_2^2}
%\end{pmatrix} \ee
Comparing explicit expressions (\ref{aaa}) with Eq. (\ref{pertpo}), we can check that:
%It can be compared to
%. Proceeding from one basis to another, we get
%\be L = \begin{pmatrix}
%0 & l_{12} \\
%l_{21} & 0
%\end{pmatrix} =
%\begin{pmatrix}
%0 & (a_1 + b_2) + i(b_1 - a_2) \\
%(a_1 - b_2) + i (b_1 + a_2) & 0
%\end{pmatrix} \ee
%and it can be checked that
\be
\frac{|l_{12}|^2}{|l_{12}|^2+|l_{21}|^2} = \frac{1}{2} + \frac{a_1 b_2 - a_2 b_1}{a_1^2 + a_2^2 + b_1^2 + b_2^2}. \label{bbb}
\ee
%The bottom-right element equals $1$ minus the top-left elements, so they also coincide.

\item At last, we need to solve the first two Eqs. (\ref{11}), (\ref{22}) and make sure that off-diagonal elements vanish in the late time limit.
Indeed, $\rho_2$ can be expressed from the first equation:
\be
\label{rho2} \rho_2 = \frac{\dot \rho_1 + 2(a_2^2 + b_2^2)\rho_1}{2(-\eps_3+a_1 a_2 + b_1 b_2)}.
\ee
After substituting this expression the second equation becomes the second-order differential equation:
\be
\ddot \rho_1 + 2(a_1^2 + a_2^2 + b_1^2 + b_2^2)\dot \rho_1 + 4\rho_1[(a_1 b_2 - a_2 b_1)^2 + \eps_3^2] = 0,
\ee
and the roots $\lambda$ are:
\be
\lambda = -(a_1^2 + a_2^2 + b_1^2 + b_2^2) \pm \sqrt{(a_1^2 + a_2^2 + b_1^2 + b_2^2)^2 - 4[(a_1 b_2 - a_2 b_1)^2 + \eps_3^2]}. \label{root}
\ee
The exponent $e^{\lambda t}$  vanishes in the limit $t\rightarrow +\infty$ for any sign of $\pm$ above, if $4[(a_1 b_2 - a_2 b_1)^2 + \eps_3^2] > 0,$ which is fulfilled due to $\eps_3 \neq 0$. Even if the square root in the r.h.s. of (\ref{root}) vanishes, i.e., $\lambda_1 = \lambda_2$, the solution $\rho_1(t)$ vanishes in the limit $t\rightarrow +\infty$, because the exponent dominates over the polynomial.
%Thus, we have proven that $\rho_1(+\infty) = 0$.
The same is true for $\rho_2(t)$, because they are related by (\ref{rho2}), and the same exponents diminish $\rho_2$ in the late time limit.
%So, $\rho_2(+\infty) = 0$.
Thus, the off-diagonal part of $\rho(t)$ vanishes asymptotically, and the complete result coincides with that obtained in (\ref{pertpo}) by means of perturbation algorithm.
\end{itemize}

%We conclude that the pointer predicted by our perturbation theory scheme is a pointer for a big variety of situations. The same pointer, according to the general solution, we %have when $H$ is an arbitrary non-degenerate $2\times 2$ matrix and $L$ is an arbitary $2\times 2$ non-zero off-diagonal matrix.
%Thus, perturbation theory scheme can be used to guess pointers for the cases, when interaction is not weak. %???????

%\newpage

\section{Conclusion}

Our first goal has been to construct FGKLS pointers, given an interaction with an environment
is weak, and perturbation theory can be applied. We have succeeded in presenting the formulas
for finding FGKLS pointers in each order of perturbation theory for non-degenerate and
degenerate Hamiltonians. We have obtained that turning an interaction with an environment
on completely changes the final states. When the system is closed, they are much more
arbitrary. If the system becomes open, its final states obey some specific sets of equations.
It means that an interaction directs the system toward a set of fixed states, pointers, that we
have been looking for throughout this work.

We have also studied particular examples of quantum harmonic oscillator with spin interacting
with external magnetic field. The first one is easy and has been solved completely. The
second one is more complicated, but we have been able to write the formulas for finding all
the orders of perturbation theory. In the third example, $H$ is an arbitrary non-degenerate $2 \times 2$
matrix and $L$ is an arbitrary $2 \times 2$ non-zero off-diagonal matrix. The pointer is predicted by
our perturbation theory scheme and is shown to coincide with the exact solution.

Several possible directions for future research can be mentioned in the context of our
study of the pointer states. The first one is the quantum cryptography where just the robust
(stationary) states, not being destroyed by external intervention, are in great need \cite{bengtsson_zyczkowski_2006}. The
second possible direction includes generalization of perturbative algorithm to the case of non-Hermitian Hamiltonians \cite{nonherm,k-2}.
The study of supersymmetry properties of open quantum
systems have been recently already started in \cite{andrianov_jan_2019} where the supersymmetric technique for
quantum engineering of systems with controllable decoherence was developed.

\section*{Acknowledgements} The research was supported by RFBR Grant No. 18-02-00264-a. The work of A.A.A. was funded by the Grant FPA2016-76005-C2-1-P and Grant 2017SGR0929 (Generalitat de Catalunya).

%\bibliography{listb}
%\bibliographystyle{ieeetr}

\end{document}